\documentclass[prp,twocolumn,superscriptaddress,floatfix,showpacs]{revtex4}
\usepackage{graphicx,amsfonts,amssymb,amsmath, hyperref}

\newif\ifhyper
\hypertrue
\ifhyper
\hypersetup{
   citecolor = {green},
   colorlinks = {true}, 
   urlcolor = {blue} 
}
\fi

\newcommand{\beq}{\begin{equation}}
\newcommand{\eeq}{\end{equation}}
\newcommand{\beqa}{\begin{eqnarray}}
\newcommand{\eeqa}{\end{eqnarray}}
\newcommand{\ket} [1] {\vert #1 \rangle}

\def\ket#1{\vert#1\rangle}

\def\Longarrow{\protect\@lra}
\def\@lra{\relbar\joinrel\relbar\joinrel\relbar\joinrel%
          \relbar\joinrel\rightarrow}

\begin{document}
\title{Disorder by disorder and flat bands in the kagome transverse field Ising model}

\author{M.~Powalski}
\affiliation{Lehrstuhl f\"ur Theoretische Physik 1, TU Dortmund, Germany}
\author{K.~Coester}
\affiliation{Lehrstuhl f\"ur Theoretische Physik 1, TU Dortmund, Germany}
\author{R.~Moessner}
\affiliation{Max Planck Institut f\"ur komplexe Systeme, D-01187 Dresden, Germany}
\author{K.~P.~Schmidt}
\email{kai.schmidt@tu-dortmund.de}
\affiliation{Lehrstuhl f\"ur Theoretische Physik 1, TU Dortmund, Germany}

\date{\rm\today}

\begin{abstract}
We study the transverse field Ising model on a kagome and a triangular lattice using high-order series expansions about the high-field limit. For the triangular lattice our results confirm a second-order quantum phase transition in the 3d XY universality class. Our findings for the kagome lattice indicate a notable instance of a disorder by disorder scenario in two dimensions. The latter follows from a combined analysis of the elementary gap in the high- and low-field limit which is shown to stay finite for all fields $h$. Furthermore, the lowest one-particle dispersion for the kagome lattice is extremely flat acquiring a dispersion only from order eight in the $1/h$ limit. This behaviour can be traced back to the existence of local modes and their breakdown which is understood intuitively via the linked cluster expansion.
\end{abstract}


\maketitle

\section{Introduction}\label{sec:introduction}
The interplay of geometric frustration and quantum fluctuations in magnetic systems provides a fascinating field of research. It gives rise to a multitude of quantum phases manifesting novel phenomena of quantum order and disorder, featuring prominent examples such as spin-ice, spin-glass and spin-liquid phases. This rich behaviour originates from the frustration of microscopically competing magnetic interactions. This can lead to a huge number of quasi-degenerate ground-state configurations and a suppression of magnetic order even at zero temperature. The macroscopic ground-state manifold of simple model systems is extremely sensitive to perturbations. Arbitrary linear combinations of the degenerate states may form novel ground states, yielding the great variety of unusual effects some of which are not fully understood to date. 
 
Perhaps the simplest and hence most tractable models featuring such behaviour are the fully-frustrated Ising systems in a transverse field \cite{liebmann1986}. On lattices like the triangular and the kagome lattice, the frustration results from the incompatibility of the antiferromagnetic bonds in a single triangular plaquette. These systems exhibit a classical ground-state manifold with a macroscopic degeneracy which is lifted in the presence of an infinitesimal transverse field inducing quantum fluctuations. 

In the limit of high transverse fields all lattices realize a non-degenerate polarized phase. The situation for weak transverse fields however is much less trivial. A large class of these frustrated transverse field Ising systems exhibits a nontrivial quantum ordered phase arising from classical disorder, a scenario which is also known as \emph{order by disorder} \cite{kano1953,villain1980,shender1982}. In contrast, the kagome lattice plays a distinguished role as it seems to be very reluctant to order \cite{moessner2000,moessner2001,nikolic2005}. It is believed to be a quantum paramagnet for arbitrary transverse fields providing a novel instance of \emph{disorder by disorder} in which quantum fluctuations select a quantum disordered state out of the classically disordered ground-state manifold. There is no conclusive evidence excluding a possible ordering of the kagome transverse field Ising model (TFIM) and its global phase diagram remains unsettled. So far, the only known Ising system featuring such an unusual disorder by disorder scenario is the one-dimensional sawtooth chain \cite{priour2001}. As a candidate for disorder by disorder in two dimensions the kagome TFIM hence is a subject of great theoretical interest.
   
The central objective of this paper is the analysis of the gapped paramagnetic phase for the kagome TFIM to check whether it indeed persists for arbitrary field values providing an instance of a disorder by disorder scenario in two dimensions. For this purpose, we investigate the energy gap by high-order series expansions about the high-field limit. Additionally, we determine the gap of the corresponding low-energy mode for an infinitesimal transverse field by means of a Lanczos based algorithm performed on appropriate finite clusters. Our findings suggest that high- and low-field gaps are indeed adiabatically connected implying that the kagome TFIM realizes a disorder by disorder scenario. 

The analoque problem on the triangular lattice is different. Here, series expansion of the one-particle gap about the high-field limit clearly indicates a second-order phase transition in the 3d XY universality class between the polarized and an ordered phase realized at low fields. Our results are in quantitative agreement with quantum Monte Carlo simulations \cite{moessner2003}. 

Finally, we show that a major difference between the triangular and the kagome TFIM lies in the existence of a localized low-energy mode for the kagome lattice that is completely flat up to order eight perturbation theory. In this order a specific momentum corresponding to the $\sqrt{3}\times\sqrt{3}$ structure is energetically favored. The emergence of this local mode and its disappearance at remarkably high orders in perturbation theory can be understood via the linked cluster expansion. Furthermore, we generalize our findings to a larger class of systems. 

This paper is organized as follows. In Sec.~\ref{sec:model} we briefly introduce the TFIM and discuss the physics of the low-field and high-field regime. In Sec.~\ref{sec:pcut} we present the derivation of the effective one-particle Hamiltonian in the high-field limit using the method of perturbative continuous unitary transformations (pCUTs) which is implemented via a linked graph expansion. The one-particle gap for the triangular and the kagome TFIM is analyzed in Sec.~\ref{sec:1QPtriangular} and in Sec.~\ref{sec:1QPkagome}. The more technical part discussing the existence of flat bands and their breakdown is placed in Sec.~\ref{sec:flatbands}. Finally, Sec.~\ref{sec:summary} provides a concluding summary.

\section{Model}
\label{sec:model} 
We study the two-dimensional Ising model in a transverse magnetic field on kagome and triangular lattices. The corresponding Hamiltonian is given by
\begin{equation}
\label{eq:ising}
	H=J \sum_{<i,j>}{\sigma^z_i \sigma^z_j}+h\sum_i{\sigma^x_i}
\end{equation}
with Pauli matrices $\sigma_i^{\alpha}$ acting on site i, the antiferromagnetic nearest-neighbour exchange $J>0$, and the strength of the transverse magnetic field $h$.

For the zero-field case $h=0$, the Hamiltonian describes a classical antiferromagnet. The energy of a single bond is minimized by an antiparallel alignment of the two neighbouring spins. However, there exists no state with all bonds satisfying the antiferromagnetic condition due to the frustrated geometry of the lattice. Consequently, a ground state is given by a configuration where every triangle has only two antiferromagnetic bonds of which there are infinitely many combinations in the thermodynamic limit. The resulting extensive ground-state degeneracy gives rise to a finite entropy density at zero temperature \cite{liebmann1986}, which is a main characteristic of highly frustrated systems. Let us also mention that the classical Ising antiferromagnet on the kagome lattice is disordered for any temperature. 

Next, consider the introduction of a small transverse field $h \ll J$ which induces quantum fluctuations by flipping the $z$-components of the spins. For any infinitesimal field the macroscopic ground-state degeneracy is lifted to some degree giving rise to a discontinuity in the entropy at zero field $h=0$ and zero temperature $T=0$. An interesting question is whether the perturbed system continues into a quantum disordered phase (disorder by disorder) or whether quantum fluctuations are selecting a quantum ordered state (order by disorder). 

We approach this question for the kagome and the triangular TFIM by considering the high-field limit ($h \gg J$) which provides an ideal setting for a perturbative approach giving also access to the low-field regime. One expects a fundamentally different behaviour of the high-field gap for an order by disorder or a disorder by disorder scenario. For the first case the elementary gap should close at the quantum critical point separating the polarized phase from the ordered phase present at low fields. In contrast, disorder by disorder implies the continuance of the disordered quantum paramagnet for arbitrary transverse fields which requires the ground state to be adiabatically connected to the low-field limit such that the high-field gap never closes.  We therefore aim at a precise determination of the excitation gap in the high-field phase for which we use a high-order series expansion.

\section{High-field expansion} 
\label{sec:pcut}
Starting from the high-field limit $h\gg J$ , we use a pCUT \cite{knetter2000,knetter2003} to derive a quasi-particle conserving effective Hamiltonian up to high order in perturbation. For this purpose, it is expedient to interpret the elementary excitations corresponding to single spin flips as quasi-particles above the vacuum which is given by the fully polarized state. The spin flips are then described in terms of hardcore bosons represented by creation and annihilation operators $b_i^{\phantom{\dagger}}$ and $ b_i^{\dagger}$. Besides the usual bosonic commutation relations these operators meet the hardcore constraint
\begin{equation}
	  b_i^{\phantom{\dagger}} b_i^{\phantom{\dagger}} = b_i^{\dagger} b_i^{\dagger} = 0 \quad 
\end{equation}
which means that every site $i$ can be occupied at most by one boson as every spin is either 'up' or 'down' (or in bosonic language: 'occupied' or 'empty').
 
\begin{figure}
	\centering
		\includegraphics[width=1.0\columnwidth]{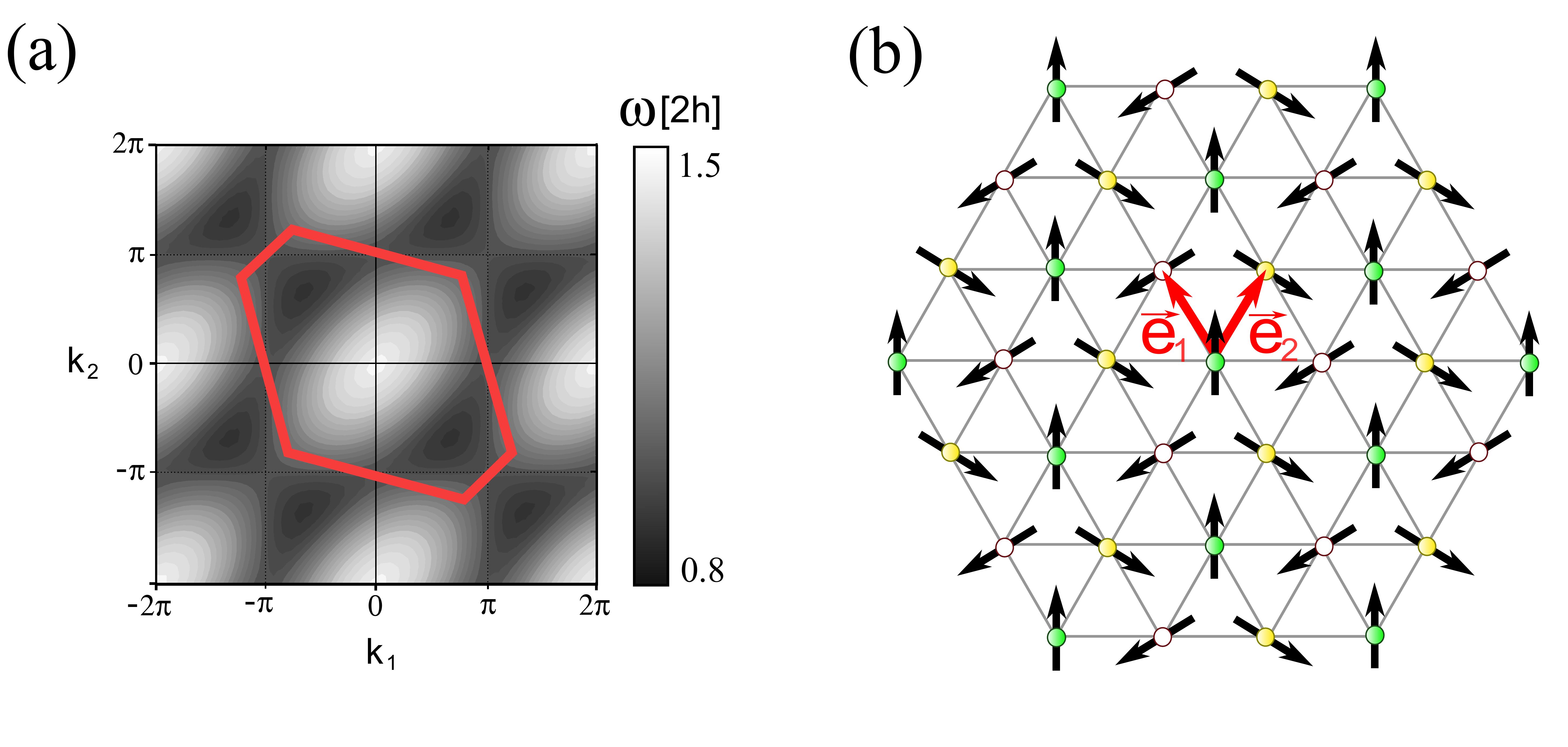}
	\caption{(a) Countour plot of the one-particle dispersion $\omega^{\rm tr}(\vec{k})$ as a function of the momentum $\vec{k}=(k_1,k_2)$ on the triangular lattice for $x=0.1$. The red (solid) line corresponds to the first Brillouin zone. (b) Illustration of the complex phase of the eigen state corresponding to the one-particle gap at $\vec{k}_{\textnormal{min}}=(\frac{2}{3}\pi, -\frac{2}{3}\pi)$. The state exhibits a $\sqrt{3}\times \sqrt{3}$-structure. Note that the structure for the second gap momentum $\vec{k}_{\textnormal{min}}=-(\frac{2}{3}\pi, -\frac{2}{3}\pi)$ is also a $\sqrt{3}\times \sqrt{3}$-structure with a different orientation. We stress that the displayed arrows do not represent spins.}
	\label{Fig:dispersion_order_triangular}
\end{figure}

The Hamiltonian (\ref{eq:ising}) can then be written as
\begin{eqnarray}
	\frac{H}{2h} &=& E_0 + \sum_i \hat{n}_i +x \sum_{\langle i,j \rangle }\left( b^\dagger_i b^{\phantom{\dagger}}_j + b^\dagger_i b^\dagger_j + {\rm h.c.}\right)   \nonumber\\
                    &=&H_0 + x \left( T_0+T_{+2}+T_{-2} \right)= H_0 + x V \label{eq:initial_hamiltonian}
\end{eqnarray}
where $x=J/2h$ is the perturbation parameter, \mbox{$E_0=-1/2$}, and $\hat{n}_i = b^\dagger_i b^{\phantom{\dagger}}_i$ is the local density on site $i$. We identify the Ising term as the perturbation $V$ which can be decomposed into a sum of $T_m$ operators incrementing (or decrementing) the number of quasi-particles by $m=\left\{+2,-2,0 \right\}$. 
The pCUT transforms the initial Hamiltonian (\ref{eq:initial_hamiltonian}) into a quasi-particle conserving effective Hamiltonian of the form
\begin{equation}
 \label{eq:eff_ham}
  H_{\textnormal{eff}}= H_0 + \sum_{n}x^n \sum_{\bar{m}=0} C(m_1, \ldots , m_n) T_{m_1}\ldots T_{m_n}  
\end{equation}
where the second sum runs over all possible $\left\{m_1,m_2,\ldots,m_n \right\}$ with $m_i =\left\{+2,-2,0 \right\}$ and $\bar{m}\equiv\sum_i m_i = 0$. The latter condition reflects the block diagonality of the effective Hamiltonian $H_{\textnormal{eff}}$ in the number of elementary excitations allowing to restrict the calculations to subspaces with a defined number of quasi-particles. The coefficients $C(m_1, \ldots , m_n)$ have been calculated as exact fractions up to high order in perturbation \cite{knetter2000}. The monomials of $T_m$ operators in order $n$ generate virtual fluctuations that involve at maximum $n$ different bonds. Due to the cluster additivity of the effective Hamiltonian, only linked virtual processes (these are processes that comprise bonds on simply connected clusters) have a finite contribution. Consequently, the matrix elements of the irreducible $n$-particle blocks of the Hamiltonian do possess a linked cluster expansion \cite{gelfand1990,gelfand1996,knetter2003}.
   
We calculated the effective Hamiltonian on the kagome (triangular) lattice in the one-quasi-particle (1QP) sector up to order $n=13$ ($n=11$) in the perturbative parameter $x$. This is done via a full linked-cluster expansion taking into accout the 1830 (1784) topologically different graphs. Note that the pCUT provides results for the thermodynamic limit that are exact up to the computed order. 

\section{Triangular lattice}
\label{sec:1QPtriangular}
In this section we discuss the one-particle properties for the triangular TFIM which is confirmed to realize an order by disorder scenario. The triangular lattice can be described by a single-site unit cell. As a consequence, the 1QP block of the effective Hamiltonian is diagonalized by Fourier transformation giving a single band for the one-particle excitations which is fully characterized by its dispersion $\omega^{\rm tr}(\vec{k})$. A contour plot of the resulting dispersion for a small coupling $x=0.1$ is depicted in Fig.~\ref{Fig:dispersion_order_triangular}(a). 
\begin{figure}
	\centering
		\includegraphics[width=\columnwidth]{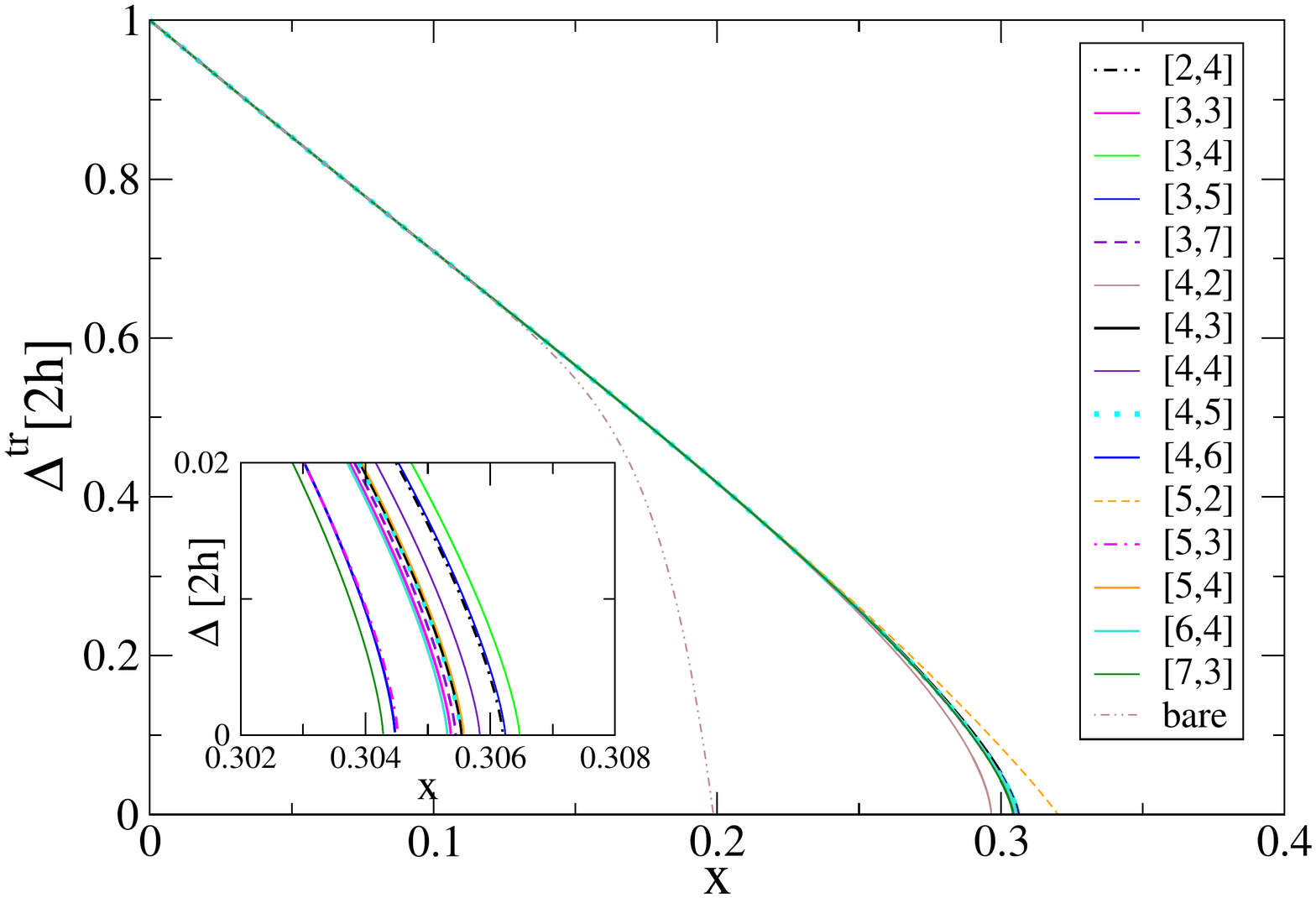}
                \includegraphics[width=\columnwidth]{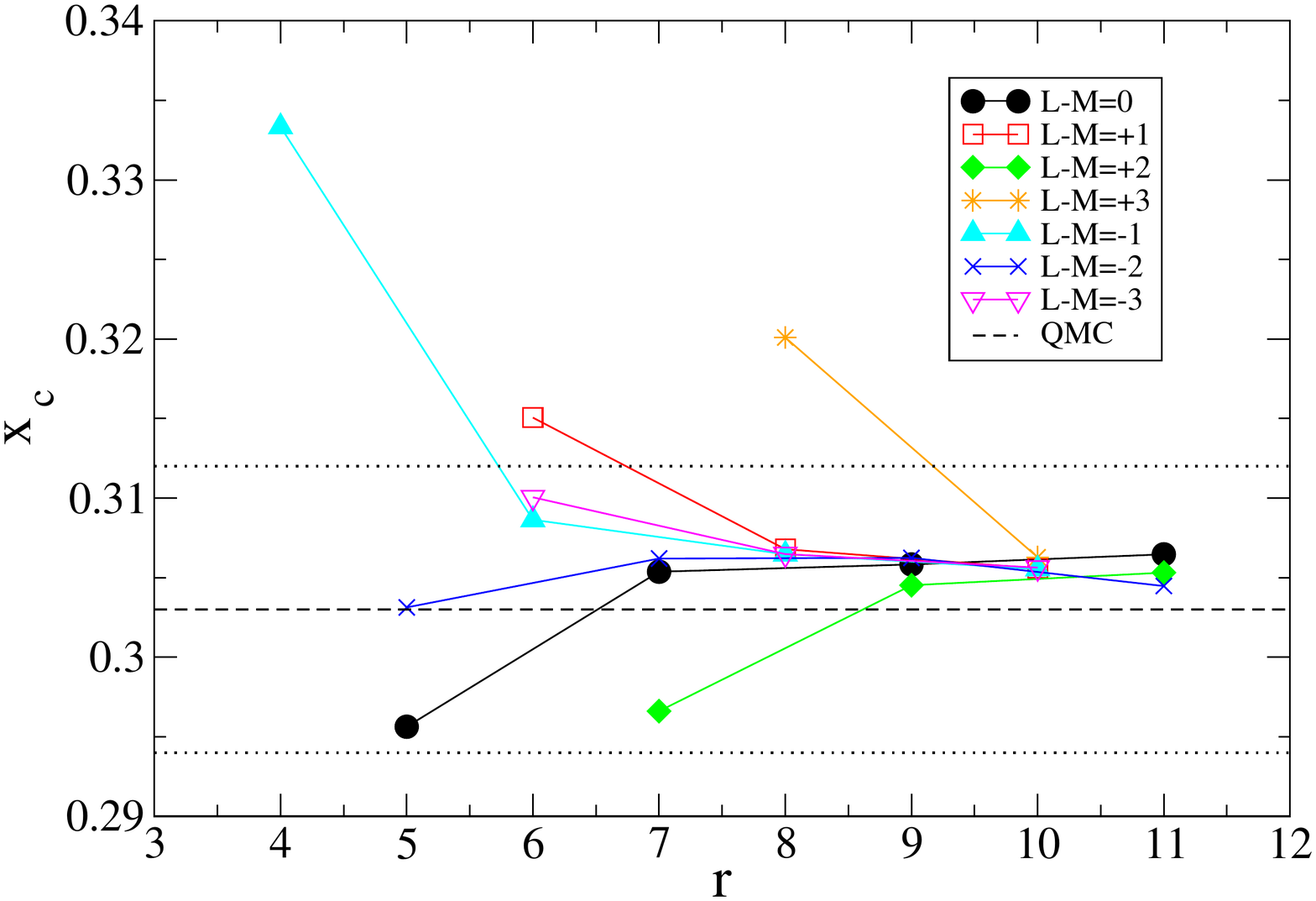}
		\includegraphics[width=\columnwidth]{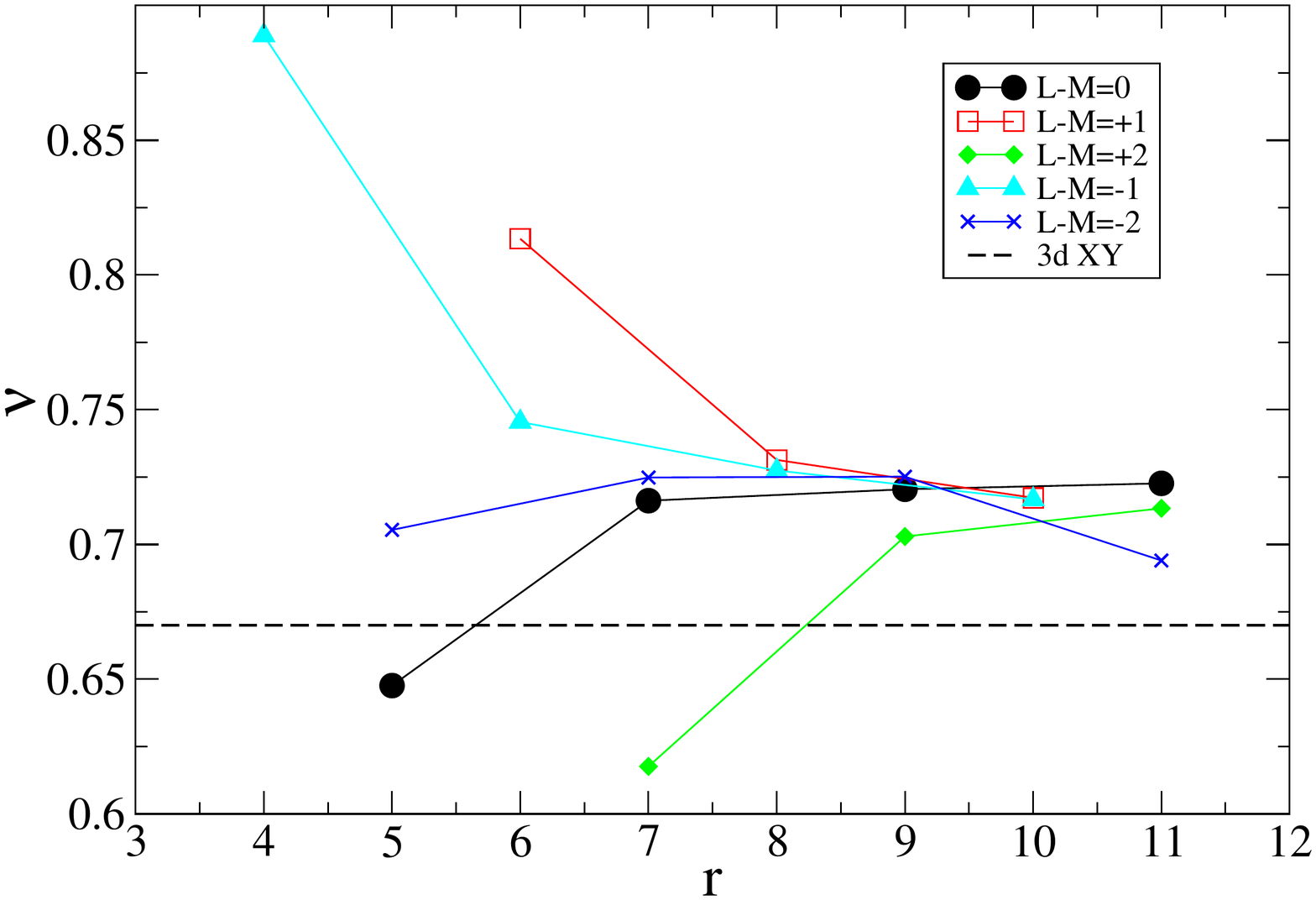}
	\caption{(a) Bare series and dlogPad\'{e} extrapolants of the one-particle gap $\Delta^{\rm tr}$ as a function of $x$ for the triangular TFIM. Inset represents a zoom close to the quantum critical point.
	(b) Illustration of the convergence behaviour of the critical values $x_{\rm c}$ resulting from the dlogPad\'e extrapolants $\left[L/M\right]$ in dependence of the considered order $r=L+M+1$. Connected points of the same color correspond to extrapolants with constant $L-M$. The dashed line represents the quantum Monte Carlo result $x_{\rm c}^\textnormal{(qmc)} = 0.303$~\cite{moessner2003}. The horizontal dotted lines indicate the corresponding margin of error $\pm 0.009$. 
	(c) Critical exponent $\nu$ extracted from dlogPad\'e extrapolants shown in (b) as a function of $r$. Dashed horizontal line indicates the critical exponent $\nu^{\rm 3dXY}\approx 0.67$ of the 3d XY universality class.}
	\label{Fig:DreieckExtrapolation}
\end{figure}
The minima of the dispersion $\omega^{\rm tr}$ are located at $\vec{k}_{\textnormal{min}}=\pm (\frac{2}{3}\pi, -\frac{2}{3}\pi)$ (see Fig.~\ref{Fig:dispersion_order_triangular}(a)) defining the one-particle gap $\Delta^{\rm tr}\equiv\omega^{\rm tr}(\vec{k}_{\textnormal{min}})$. Up to order $n=11$, this gap is given by
\begin{eqnarray}
	\Delta^{\rm tr}= 1 &-& 3 x + \frac{3}{2} x^2 - \frac{15}{2} x^3 + \frac{243}{8} x^{4} - \frac{1671}{8}x^5 \nonumber\\ &+& \frac{22275}{16} x^6 - \frac{162855}{16}x^7 + \frac{9700617}{128} x^8 \nonumber\\ &-& \frac{595490847}{1024} x^9 + \frac{2010551941313}{442368} x^{10} \nonumber\\ &-& \frac{1777064899901}{49152}x^{11} \quad.   \label{eq:triangulargap}
\end{eqnarray}
It is instructive to plot the complex phase of the eigen state corresponding to the gap momenta $\vec{k}_{\rm min}$ which is shown in Fig.~\ref{Fig:dispersion_order_triangular}(b). For the triangular lattice a $\sqrt{3}\times\sqrt{3}$ structure is found and one expects that linear combinations of such structures are the most preferable ordering patterns (see below).
 
In order to detect a possible second-order quantum phase transition, we use dlogPad\'e techniques to extrapolate the one-particle gap $\Delta^{\rm tr}$. To this end, various extrapolants $\left[L,M\right]$ are constructed where $L$ denotes the order of the numerator and $M$ the order of the denominator. In Fig.~\ref{Fig:DreieckExtrapolation} the corresponding results for the one-particle gap $\Delta^{\rm tr}$ are displayed.

The gap shows a strong tendency to close in a narrow window around $x_{\rm c}\approx 0.305$. These results are in quantitative agreement with quantum Monte Carlo (QMC) simulations giving \mbox{$x_{\rm c}^\textnormal{(qmc)} = 0.303\pm0.009$} \cite{moessner2003}. The convergence behaviour of the critical values $x_{\rm c}$ for different extrapolants compared to $x_{\rm c}^\textnormal{(qmc)}$ is shown in Fig.~\ref{Fig:DreieckExtrapolation}(b). Altogether, our findings suggest a second-order quantum phase transition in the vicinity of \mbox{$x_{\rm c}^\textnormal{(qmc)}=0.303$} and $x_{\rm c}=0.305$.

Furthermore, one can extract the critical exponent $\nu$ from the dlogPad\'e extrapolation. The corresponding results are shown in Fig.~\ref{Fig:DreieckExtrapolation}(c). Averaging over all dlogPad\'e extrapolants $[n,m]$ of order 11 with $|n|,|m|\geq 2$ , one obtains 
\mbox{$\nu=0.708\pm0.012$} which is slightly larger than the expected critical exponent $z\nu\approx0.67$ for the 3d XY universality class \cite{hasenbusch1994}. 
Interestingly, if one biases the dlogPad\'e extrapolants to have a pole at $x_{\rm c}^\textnormal{(qmc)}=0.303$, the corresponding averaged exponent \mbox{$\nu^{\rm bias}=0.658\pm 0.016$} is much closer to 3d XY. 

The closing of the gap corresponds to a quantum phase transition between an disordered paramagnetic phase for $x<x_{\rm c}$ and a quantum ordered phase for $x>x_{\rm c}$. The symmetry of the ordered phase is reduced according to the $\sqrt{3}\times \sqrt{3}$ structure which is consistent with the bond ordered phase suggested in Ref.~\onlinecite{moessner2000}. 

The triangular TFIM therefore realizes an order by disorder scenario. A second-order quantum phase transition in the 3d XY universality class is detected by the closing of the high-field gap. This is different for the analoque problem on the kagome lattice as we will elaborate on in the next section.
   
\section{Kagome lattice}
\label{sec:1QPkagome}
\begin{figure}
	\centering
		\includegraphics[width=1.0\columnwidth]{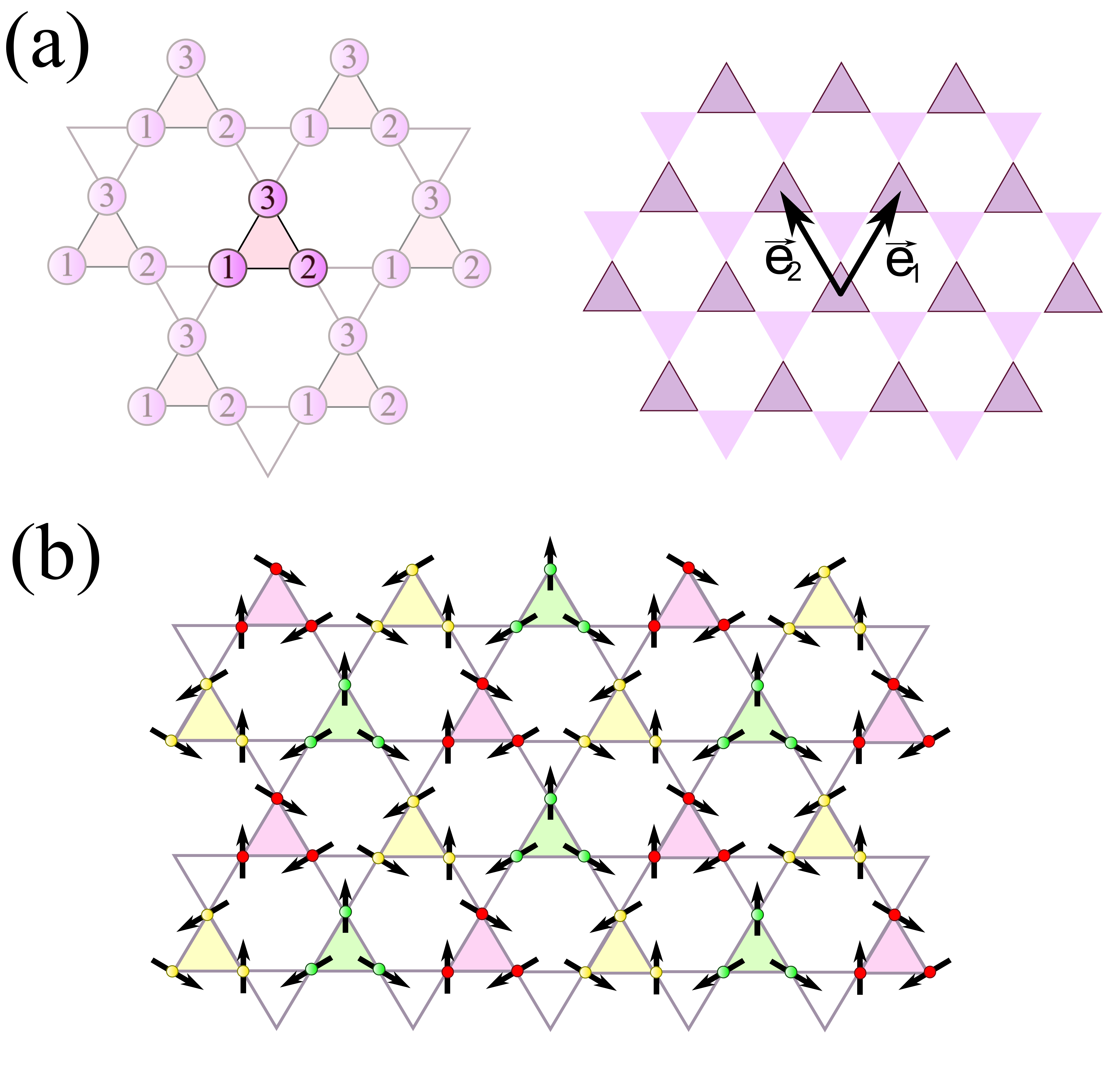}
	\caption{(a) The kagome lattice has a three-site unit cell. We define unit vectors $\vec{e}_1$ and $\vec{e}_2$ as shown on the right-hand side of the upper panel. The position of a unit cell is then given by $(n_1, n_2)$ with $n_i\in\mathcal{Z}$. The three sites of the unit cell are denoted by $\{ 1,2,3 \}$.
	(b) Illustration of the eigen state corresponding to the one-particle gap. Arrows denote the complex phase of this eigenvector. The pattern realizes a $\sqrt{3}\times \sqrt{3}$-structure. Note that the structure for the second gap momentum $\vec{k}_{\textnormal{min}}=-(\frac{2}{3}\pi, -\frac{2}{3}\pi)$ is also a $\sqrt{3}\times \sqrt{3}$-structure with a different orientation. We stress that the displayed arrows do not represent spins.}
	\label{Fig:Kagome_order}
\end{figure}

In this section we aim at determining the one-particle gap of the kagome TFIM on the whole parameter axis. Since the kagome lattice has a three-site unit cell (see Fig.~\ref{Fig:Kagome_order}(a)), the effective one-particle Hamiltonian $H_{\rm eff}^{\rm 1QP}$ is only reduced to an hermitian $3\times3$ matrix $\omega_{lm}(k_1,k_2)$ with $l,m\in\{1,2,3\}$ per momentum $\vec{k}$ using Fourier transformation. One therefore has three one-particle bands for the kagome TFIM. Due to the lattice symmetries, it is sufficient to calculate just the two matrix elements $\omega_{11}(k_1,k_2)$ and $\omega_{12}(k_1,k_2)$. The remaining matrix elements can be obtained by the following relations
\begin{eqnarray}
	  \omega_{22}(k_1,k_2)&=&\omega_{11}(k_2,k_1)\nonumber\\
	  \omega_{33}(k_1,k_2)&=&\omega_{11}(-k_2,k_1-k_2)\nonumber\\
	  \omega_{13}(k_1,k_2)&=&\omega_{12}(k_1-k_2,-k_2)\nonumber\\
	  \omega_{23}(k_1,k_2)&=&\omega_{12}(k_2-k_1,-k_1)\nonumber\\
	  \omega_{lm}&=&(\omega_{ml})^* \quad .
\end{eqnarray}
There are two dispersive high-energy bands exhibiting a finite bandwidth already in first-order perturbation theory. Most interestingly, the remaining low-energy band $\omega_{\textnormal{low}}$ is completely flat up to and including order $n=7$
\begin{equation}
	 \omega_{\textnormal{low}}^{(7)}(\vec{k})=1-2x+2x^2-3x^5+\frac{243}{8}x^6-\frac{2609}{8}x^7 \quad .
\end{equation}

\begin{figure}[htbp]
	\centering
		\includegraphics[width=0.7\columnwidth]{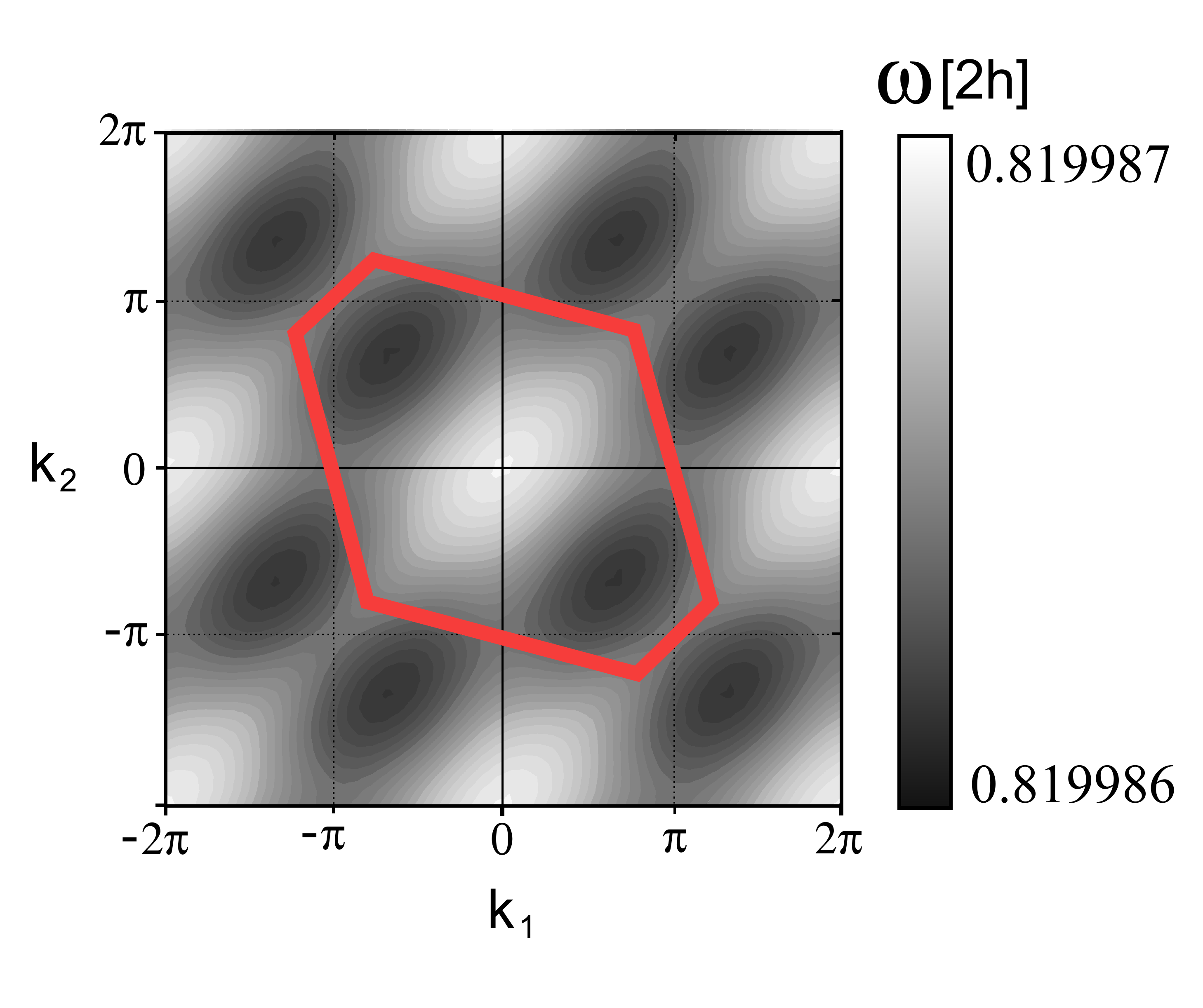}
	\caption{Lowest one-particle energy band $\omega^{(13)}_{\rm low}(\vec{k})$ as a function of $\vec{k}=(k_1,k_2)$ for the kagome TFIM at $x=0.1$. Red solid line indicates the first Brillouin zone. Note the extreme smallness of the bandwidth.}
	\label{Fig:kagome_dispersion_low}
\end{figure}

Note that in general the presence of such a flat one-particle band indicates the existence of a local mode. We give a detailed discussion on the physics of such local modes in Sec.~\ref{sec:flatbands}. Here we continue with the analysis of the one-particle gap. As the low-energy band starts to disperse in order $n=8$, quantum fluctuations occuring in this order lift the degeneracy in momentum-space and select a specific gap momentum which is energetically favored. The resulting dispersion $\omega^{(13)}_{\rm low}(\vec{k})$ of order 13 is shown for $x=0.1$ in Fig.~\ref{Fig:kagome_dispersion_low}.

The locations of the energy gap are $\vec{k}_{\textnormal{min}}=\pm (\frac{2}{3}\pi, -\frac{2}{3}\pi)$. In order $n=13$, one finds the following expression for the one-particle gap
\begin{eqnarray}
 \Delta &=&1-2x+2x^2-3x^5+\frac{243}{8}x^6-\frac{2609}{8}x^7 \nonumber\\
        && +\frac{28531}{16}x^8-\frac{3448807}{512}x^9+\frac{149125883}{6144}x^{10}\nonumber\\
        &&-\frac{1340557475}{12288}x^{11}+\frac{168486121055}{294912}x^{12}\nonumber\\
        &&-\frac{113062814254323109}{36238786560}x^{13}\, .
\end{eqnarray}

 The form of the resulting eigenvector (up to a global phase shift) is given by
\begin{equation}
	\vec{v} = \frac{1}{\sqrt{3}}\left(1, e^{i\frac{2}{3}\pi}, e^{-i\frac{2}{3}\pi}\right) \quad ,
\end{equation}
where the three coordinates of $\vec{v}$ correspond to the three sites of the unit cell (see Fig.~\ref{Fig:Kagome_order}). If one plots the complex phase of the full lattice eigen state, one finds a $\sqrt{3}\times\sqrt{3}$-structure illustrated in Fig.~\ref{Fig:Kagome_order}(b). In an analogue fashion to the triangular lattice, one would expect that, if the kagome TFIM realizes an order by disorder scenario, the ordered phase has a $\sqrt{3}\times \sqrt{3}$ structure. Interestingly, this is suggested as the ordering pattern in the bond-ordered phase of the kagome Ising antiferromagnet in a longitudinal field \cite{moessner2001}. 

Furthermore, a high-temperature expansion of the classical kagome Heisenberg model displays a striking similarity to our findings: Fluctuations (in this case thermal) up to order $n=7$ in $\beta=1/T$ do not pick any specific momentum. It is only in order eight that the momentum corresponding to a $\sqrt{3}\times \sqrt{3}$-structure is favored \cite{harris1992,reimers1993}. Consequently, thermal fluctuations of the classical kagome Ising model behave not unlike quantum fluctuations in the kagome TFIM in the high-field limit. One might wonder whether the origin of this equivalence can be traced back to geometrical relations. This is indeed true as we will discuss in Sec.~\ref{sec:flatbands}.  

In the following we analyze the one-particle gap on the whole field axis in order to clarify whether the kagome TFIM realizes a disorder by disorder scenario.  At this point we assume that binding effects are small for the TFIM and, consequently, multi-particle sectors play no role for the excitation gap (for any $J/h$). This assumption is reasonable since the leading contribution to the two-particle interaction is repulsive and of order $n=2$ perturbation theory.

\begin{figure}[htbp]
	\centering
		\includegraphics[width=1.0\columnwidth]{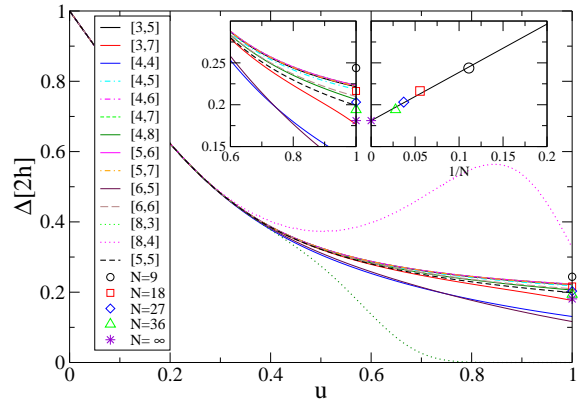}
	\caption{One-particle gap $\Delta/2h$ as a function of $u$. Lines correspond to dlogPad\'e extrapolants in $u=\frac{x}{x+1}$ while symbols represent the first-order degenerate perturbation theory results of the low-field limit for the different finite-size samples. ({\it Left Inset}): Zoom around $u=1$. ({\it Right Inset}): First-order degenerate perturbation theory results of the one-particle gap against $1/N$. Solid black line represents a linear fit of the data.}
	\label{Fig:Kagome_u_extrapolation}
\end{figure}

For the extrapolation it is convenient to perform an Euler transformation 
\begin{equation}
	x =: \frac{u}{u-1}
\end{equation}
which maps the parameter axis $x\in {[0,\infty[} $ on the finite interval $u\in [0,1]$. One expects that the dlogPad\'e extrapolants in $u$ are more stable within the finite interval $u\in{[0,1]}$ allowing to estimate the gap for $x\rightarrow \infty$ which corresponds to $u=1$. The resulting extrapolations of the high-field gap are shown in Fig.~\ref{Fig:Kagome_u_extrapolation}.

For $u<0.4$, all extrapolations as well as the bare series are well converged. A closing of the gap in this region can therefore be excluded. Note that the bare series of $\Delta(x)$ converges very slowly as the absolute value of the coefficients grows exponentially with ascending order and, additionally, as the sign of the coefficients alternate. But also far beyond the high-field regime most of the extrapolants indicate a clear trend. Though isolated extrapolants exhibit instabilities, there is no tendency that the gap closes at a certain critical value. Moreover, most of the extrapolations approach a finite value $\Delta_{\infty}\approx 0.20 \pm 0.03$ for $u=1$. 

From the pCUT result the lowest one-particle mode in the high-field limit is characterized by a momentum $\vec{k}_{\textnormal{gap}}=(\frac{2}{3}\pi,-\frac{2}{3}\pi)$ whereas the ground state has $\vec{k}=(0,0)$. If the gap stays finite for arbitrary $x$ with $\Delta(x\rightarrow \infty)\geq0$, corresponding to a disorder by disorder scenario, it implies that the ground state and the lowest one-particle excitation in the high-field limit can be adiabatically connected to the low-field regime. It would be therefore desirable to  additionally analyze the low-field limit of the kagome TFIM which we focus on next.

In the low-field limit, the kagome TFIM is best expressed as follows
\begin{equation}
	 \frac{H_{\textnormal{TFIM}}}{J} = \sum_{\langle i,j \rangle} \sigma_i^z \sigma_j^z + \frac{h}{J}\sum_{i} \sigma_i^x= H_0 + yV  \quad,
\end{equation}
where $y=h/J$. For zero field $y=0$, the Hamiltonian describes the classical kagome Ising antiferromagnet with macroscopic ground-state degeneracy. An infinitesimal field $y\ll 1$ lifts this degeneracy leading to an energy splitting between states with different momenta $\vec{k}$. 

We assume that the two lowest energy levels are characterized by momenta $\vec{k}=0$ (ground state) and $\vec{k}_{\textnormal{gap}}$ (lowest excited state) such that they are adiabatically connected to the corresponding states in the high-field limit. One can show that the energy gap in the low-field limit $\tilde{\Delta}(y)$ (in units of $J$) and the gap $\Delta(x)$ (in units of $2h$) are related by the following equation
\begin{equation}
	\frac{\partial}{\partial y} \tilde{\Delta}(y)\big\vert_{y=0} = 2\lim\limits_{x \rightarrow \infty}{\Delta(x)}\quad. \label{gaprelation}
\end{equation}
Consequently, the gap $\Delta(x\rightarrow \infty)$ can be determined by means of first-order degenerate perturbation theory about the low-field regime. 

Due to the extensive ground-state degeneracy, one has already in order-one degenerate perturbation theory infinitely many states in the thermodynamic limit. We therefore approach this problem by applying first-order degenerate perturbation theory on appropriate finite clusters of different system sizes in order to obtain an estimate of the energy splitting in the thermodynamic limit by finite-size scaling. One expects that the energy splitting converges to the gap $\Delta(x\rightarrow \infty)$ rather rapidly with increasing system size because a finite energy gap implies a finite correlation length.

\begin{figure}[htbp]
	\centering
		\includegraphics[width=1.0\columnwidth]{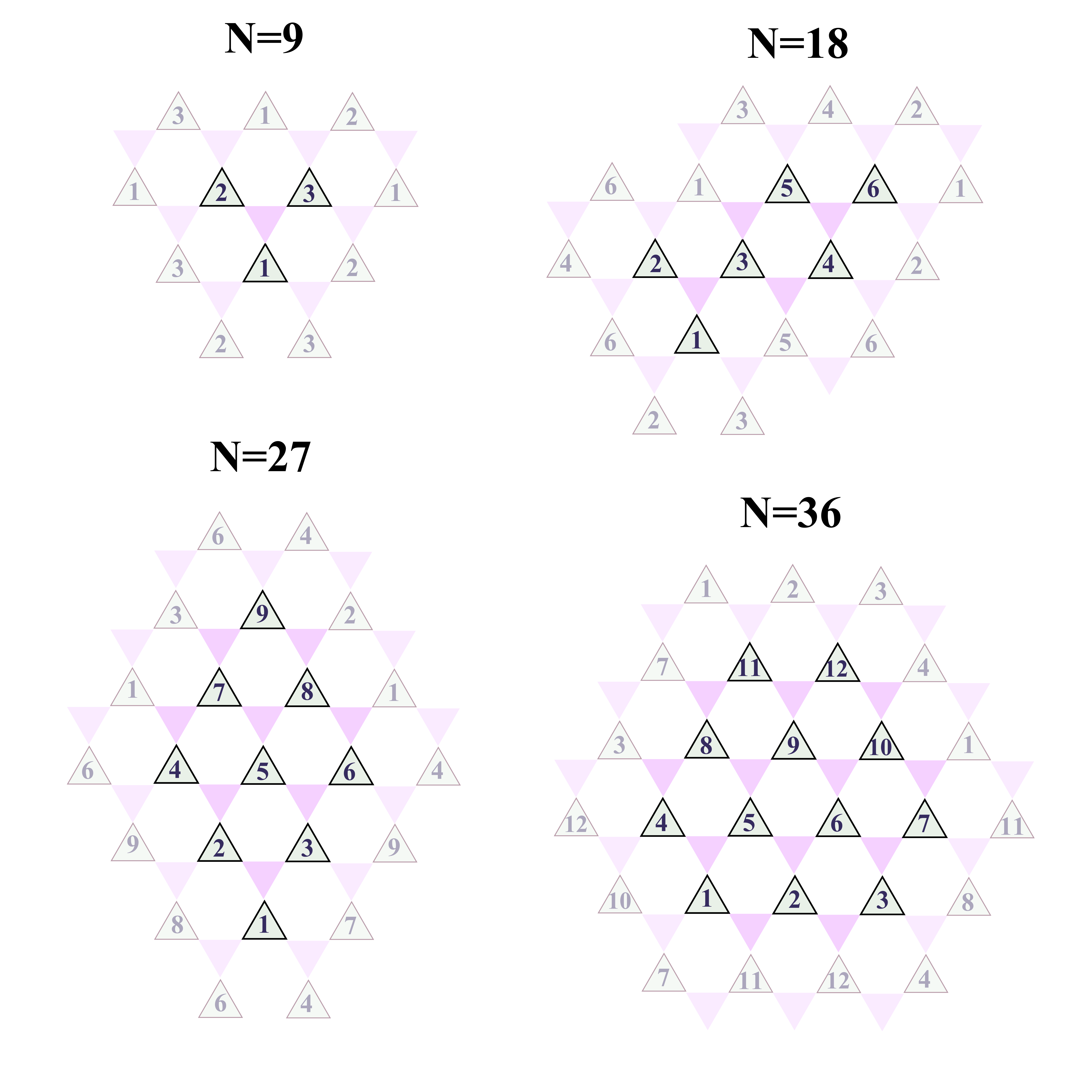}
	\caption{The finite-size samples of the kagome lattice with periodic boundary conditions which are used for the first-order degenerate perturbation theory in the low-field limit. Here $N$ denotes the total number of spins. The finite-size samples are built by the thick black triangles labeled by an integer number. Thin triangles are displayed to visualize periodic boundary conditions.}
	\label{Fig:finite_systems}
\end{figure}

In order to ensure conservation of momentum the clusters have to exhibit translational invariance which can be realized by periodic boundary conditions. Since finite systems only allow for discrete values of $\vec{k}$ depending on their sizes and geometrical shapes, the choice of the clusters plays a crucial role. We are interested in the energy having the gap momentum $\vec{k}_{gap}$ which is realized in systems allowing for a $\sqrt{3}\times\sqrt{3}$-structure. Their corresponding unit cell contains $N=9$ spins. Fig.~\ref{Fig:finite_systems} shows the appropriate clusters considered in this paper. 

Technically, we use a Lanczos based algorithm which allows to calculate the lowest first-order energy corrections in subspaces with fixed momentum $\vec{k}$. For the case of $N=36$ spins (ground-state degeneracy $g\sim 10^{10}$) we need about $n_{L}\approx100$ Lanczos steps for an accuracy of $10^{-6}$.  

The obtained energy gaps for the different system sizes are displayed in the inset of Fig.~\ref{Fig:Kagome_u_extrapolation}. Using a $1/N$-scaling, one finds for the thermodynamic limit the estimate $\Delta_{x=\infty}/2h \approx 0.18$.  This value is remarkably consistent with the value obtained from the high-field expansion.

\begin{figure}[htbp]
	\centering
		\includegraphics[width=1.1\columnwidth]{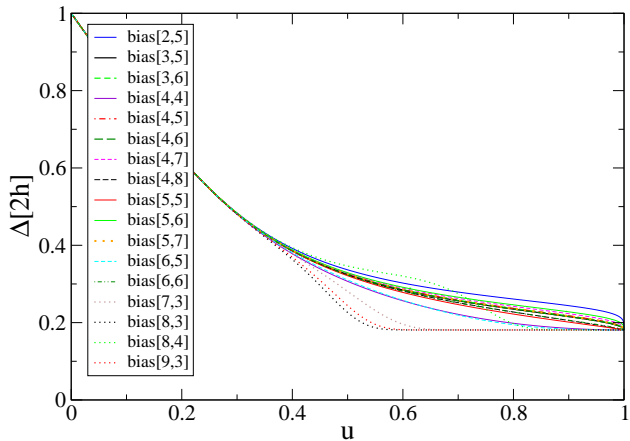}
	\caption{Biased dlogPad\'e extrapolations of the one-particle gap as a function of $u$. The extrapolants are biased to \mbox{$\Delta(u=1)/2h = 0.18$} obtained from the $1/N$-scaling in the low-field limit.}
	\label{Fig:Kagome_u_biased_etrapolation}
\end{figure}

As a further consistency check, the value of the high-field extrapolants for $u=1$ is biased to the value \mbox{$\Delta_{\infty}/2h=0.18$} calculated in the low-field limit. The resulting extrapolants are shown in Fig.~\ref{Fig:Kagome_u_biased_etrapolation}. The enhanced stability of the extrapolation indicates a reasonable consistency between the high-field and the low-field results. Altogether, the results provide considerable evidence that the gap stays finite for all values of $x$. We conclude that the kagome TFIM exhibits a disordered paramagnetic phase on the whole parameter axis. 

\section{Flat bands}
\label{sec:flatbands}
One of the key differences between the kagome and the triangular TFIM is the existence of a completely flat low-energy band up to order eight perturbation theory for the kagome lattice. It is therefore the purpose of this section to analyze the physics of such flat bands in more detail. This is done first for the kagome TFIM and it is later generalized to a larger class of systems. 

In general, the momentum independence of an one-particle energy band indicates the existence of a localized one-particle mode as any superposition of momentum basis states provides an eigen state of $H_{\rm eff}^{\rm 1QP}$. A uniform superposition in momentum space corresponds to a state that is completely localized in the spatial domain. The effective Hamiltonian $H_{\rm eff}^{\rm 1QP}$ of the kagome TFIM therefore gives rise to a local one-particle mode which is an exact eigen state of the Hamiltonian up to order $7$ in $x$. As the low-energy mode starts to disperse in order $n=8$, quantum fluctuations occuring in this order lift the degeneracy in momentum space and local modes are destroyed in this order for the kagome TFIM. 

\begin{figure}
	\centering
		\includegraphics[width=1.00\columnwidth]{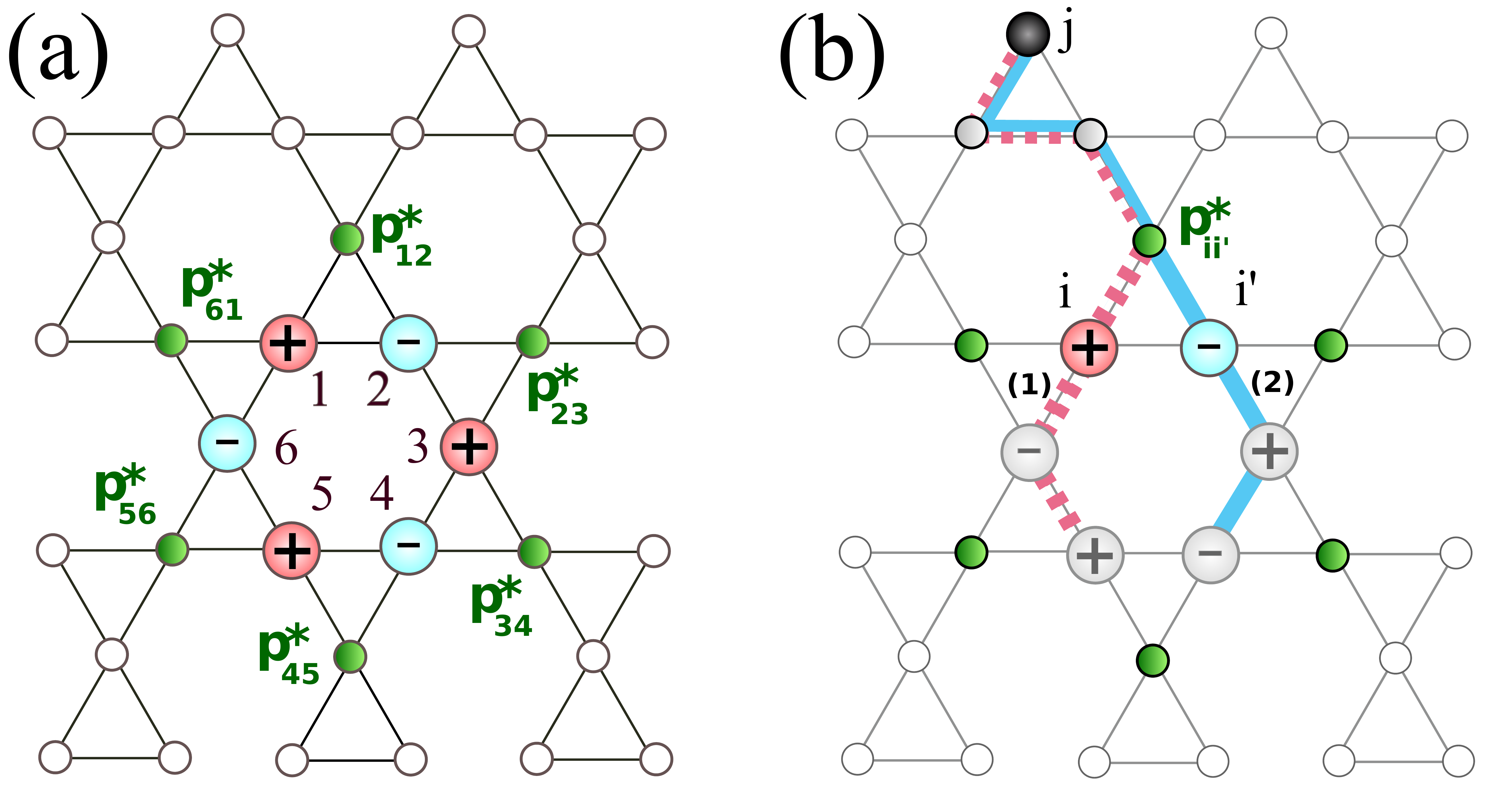}
	\caption{(a) Localized mode $|\psi_{\rm loc}\rangle$ on a hexagon of the kagome lattice. The notion $\pm$ denotes the sign in front of the one-particle states contained in $|\psi_{\rm loc}\rangle$. Filled dark circles correspond to the six sites $p^*_{ij}$ surrounding the central hexagon. (b) Example of a destructive interference between the two virtual hopping processes $(1)$ (thick dashed bonds) and $(2)$ (thick solid bonds) exhibiting the same absolute value $t_{ij}^{(1)}=t_{i'j}^{(2)}$ of the hopping amplitude. The initial states are given by the one-particle states $|i\rangle$ and $|i'\rangle$ while the final state of both processes is given by $|j\rangle$. Due to the opposite prefactor of  $|i\rangle$ and $|i'\rangle$ in $|\psi_{\rm loc}\rangle$, both processes interfere destructively.}
	\label{Fig:kagome_local}
\end{figure}

The localized mode of the kagome TFIM is given by a superposition of one-particle states located on a single hexagon with alternating relative phases as shown in Fig.~\ref{Fig:kagome_local}(a). According to the displayed numbering a localized mode on a certain hexagon can be written as 
\begin{equation}
	\ket{\psi_{\textnormal{loc}}} = \frac{1}{\sqrt{6}}\left(\ket{1} - \ket{2}+\ket{3}-\ket{4}+\ket{5}-\ket{6}\right)
\end{equation}
where $\ket{i}$ denotes a one-particle state with a particle (spin flip) located at site $i$. 

Based on the quasi-particle picture the locality can be explained
by destructive intereferences of virtual hopping processes which prevents the particles
from escaping the hexagon. 

We start by explicitly discussing the locality of $|\psi_{\rm loc}\rangle$ in leading order perturbation theory. The first-order contribution $H_{\rm eff}^{(1)}$  of the effective Hamiltonian (\ref{eq:eff_ham}) is given by
\begin{equation}
 H_{\rm eff}^{(1)}=x T_0 = x\sum_{b} \tau_{0}^{b} \quad ,
\end{equation}
where the sum is taken over all bonds of the lattice and the $\tau_{0}^{b}$ correspond to nearest-neighbour hopping processes on bond $b$. The state $|\psi_{\rm loc}\rangle$ is then obvisously an eigen state of $H_{\rm eff}^{(1)}$, since hopping processes escaping the hexagon interfere pairwise destructively on each of the six sites $p^*_{ij}$ surrounding the hexagon (see also Fig.~\ref{Fig:kagome_local}a). Such destructive interferences are the guiding principle to understand the locality of $|\psi_{\rm loc}\rangle$ also in higher orders, which is best understood via the linked cluster expansion.

Indeed, due to the linked cluster theorem, the general expression of the effective Hamiltonian in order $n$ can be decomposed as a sum of linked virtual hopping processes as follows
 \begin{eqnarray}
\label{eq:heff_n}
 H_{\rm eff}^{(n)} &=& x^n \sum_{\bar{m}=0} C(\bar{m}) \sum_{b_1,b_2,\ldots,b_n} \tau_{m_1}^{b_1}\tau_{m_2}^{b_2}\ldots \tau_{m_n}^{b_n} \nonumber\\
               &=& x^n \sum_{C_{\rm sub}}\, \sum_{ \substack{C_{\rm sub}}=\bigcup_{i=1}^n \{b_i\}}\sum_{\bar{m}=0} C(\bar{m})\,\tau_{m_1}^{b_1}\tau_{m_2}^{b_2}\ldots \tau_{m_n}^{b_n} \nonumber\\
               &=& x^n \sum_{C_{\rm sub}} \Gamma_{C_{\rm sub}}^{(n)}\quad ,
\end{eqnarray}
where the last sum is taken over all connected subclusters $C_{\rm sub}$ built by the collection of linked bonds $\{ b_i \}$. Note that the same bond can appear several times in the same operator product $\tau_{m_1}^{b_1}\tau_{m_2}^{b_2}\ldots \tau_{m_n}^{b_n}$ where $\tau_{m_j}^{b_i}$ acts on bond $b_i$ and $m_j\in\{2,-2,0\}$ refers to a pair creation, pair annihilation, or single-particle hopping process. The operator $\Gamma_{C_{\rm sub}}^{(n)}$ represents all virtual fluctuations of the connected subcluster $C_{\rm sub}$ in order $n$. 

Here we are interested in the single-particle hopping amplitudes $t_{ij}$ describing a hopping from site $i$ to site $j$
\begin{equation}   
  t_{ij} = \langle j| H_{\rm eff}| i\rangle - \delta_{i,j} \langle 0| H_{\rm eff}| 0\rangle \quad ,
\end{equation}
where $| 0\rangle$ denotes the zero-particle state being equivalent to the fully polarized state. Consequently, the contribution of a connected subclusters $C_{\rm sub}$ to such a hopping element is given by 
\begin{eqnarray}  
t_{ij}^{C_{\rm sub}}  &=&  \sum_n\left(  \langle j| \Gamma_{C_{\rm sub}}^{(n)} | i\rangle - \delta_{i,j} \langle 0| \Gamma_{C_{\rm sub}}^{(n)} | 0\rangle \right) \nonumber\\
                  &=&  \sum_n t_{ij}^{C_{\rm sub},(n)} \quad . 
\end{eqnarray}
In the following we visualize a contribution $t_{ij}^{C_{\rm sub}}$ by coloring the bonds $b_i$ of the subcluster $C_{\rm sub}$. Most importantly, the $t_{ij}^{C_{\rm sub}}$ only depend on the topology of the subcluster $C_{\rm sub}$. 
 
The latter property is crucial to pinpoint the properties of the local mode $|\psi_{\rm loc}\rangle$. The dispersion emerging in order $n = 8$ is then related to non-trivial processes that do not occur for $n < 8$. In essence, all processes up to order $n=7$ cancel out. This can be understood as follows.

\begin{figure}
	\centering
		\includegraphics[width=1.0\columnwidth]{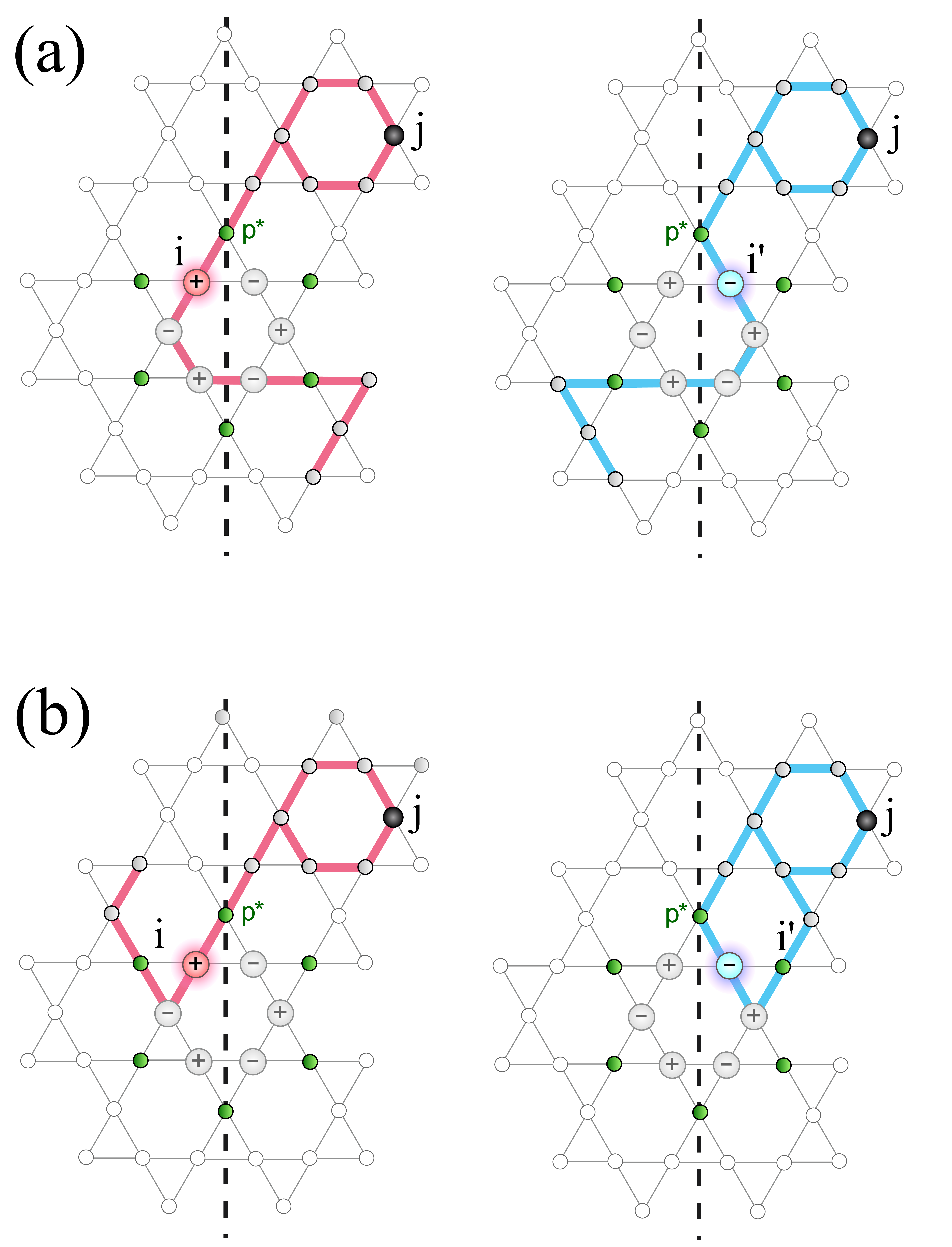}
	\caption{(a) Left figure illustrates a hopping process $t_{ij}^{C_{\rm sub}}$ on a subcluster $C_{\rm sub}$ visualized as thick solid bonds. The right part represents the counter process $t_{i'j}^{\tilde{C}_{\rm sub}}$ on $\tilde{C}_{\rm sub}$. The dashed vertical line is the symmetry axis through site $p^*$ when acting on $|\psi_{\rm loc} \rangle$.
	(b) Example of a hopping process that is not annihilated by a counter process. Any counter cluster has a different topology and therefore a different contribution. This leads to a finite hopping amplitude of a particle leaving the hexagon.}
	\label{Fig:countertrafo}
\end{figure}

In order to escape the hexagon any particle has to traverse at least one of the sites $p^*_{ij}$ enclosing the hexagon. Now, envision a process $t_{ij}^{(1)}$ on a subcluster (1) as shown in Fig.~\ref{Fig:kagome_local}(b). The initial state $|i\rangle$ is assumed to have a prefactor $+1$ in $|\psi_{\rm loc}\rangle$ and one considers the particle to escape through the site $p^*_{ii'}$ to the final state $|j\rangle$. A possible counter process on a subcluster (2) is then given by mirroring only the part of subcluster (1) being "below" the site $p^*_{ii'}$ (one can always perform a global rotation of the lattice such that $p^*_{ij}$ is located on top of the hexagon). In this case the topology of the subcluster (2) is equivalent to the subcluster (1) and, as a consequence, one has $t_{ij}^{(1)}=t_{i'j}^{(2)}$. Since the prefactor in $|\psi_{\rm loc}\rangle$ of the state $|i'\rangle$ is $-1$, the sum of both processes contained in Eq.~\ref{eq:heff_n} cancel out when acting on the state $|\psi_{\rm loc}\rangle$ (see Fig.~\ref{Fig:kagome_local}(b)). A second more complex example of such counter clusters is shown in Fig.~\ref{Fig:countertrafo}(a).

There are also subclusters that do not possess counter clusters of the same topology as exemplified in Fig.~\ref{Fig:countertrafo}(b). These subclusters contain virtual fluctuations that are not completely annihilated by counter processes when acting on $|\psi_{\rm loc} \rangle$. Hence, they correspond to fluctuations that lead to a breakdown of the local mode.

\begin{figure}[htbp]
	\centering
		\includegraphics[width=1.0\columnwidth]{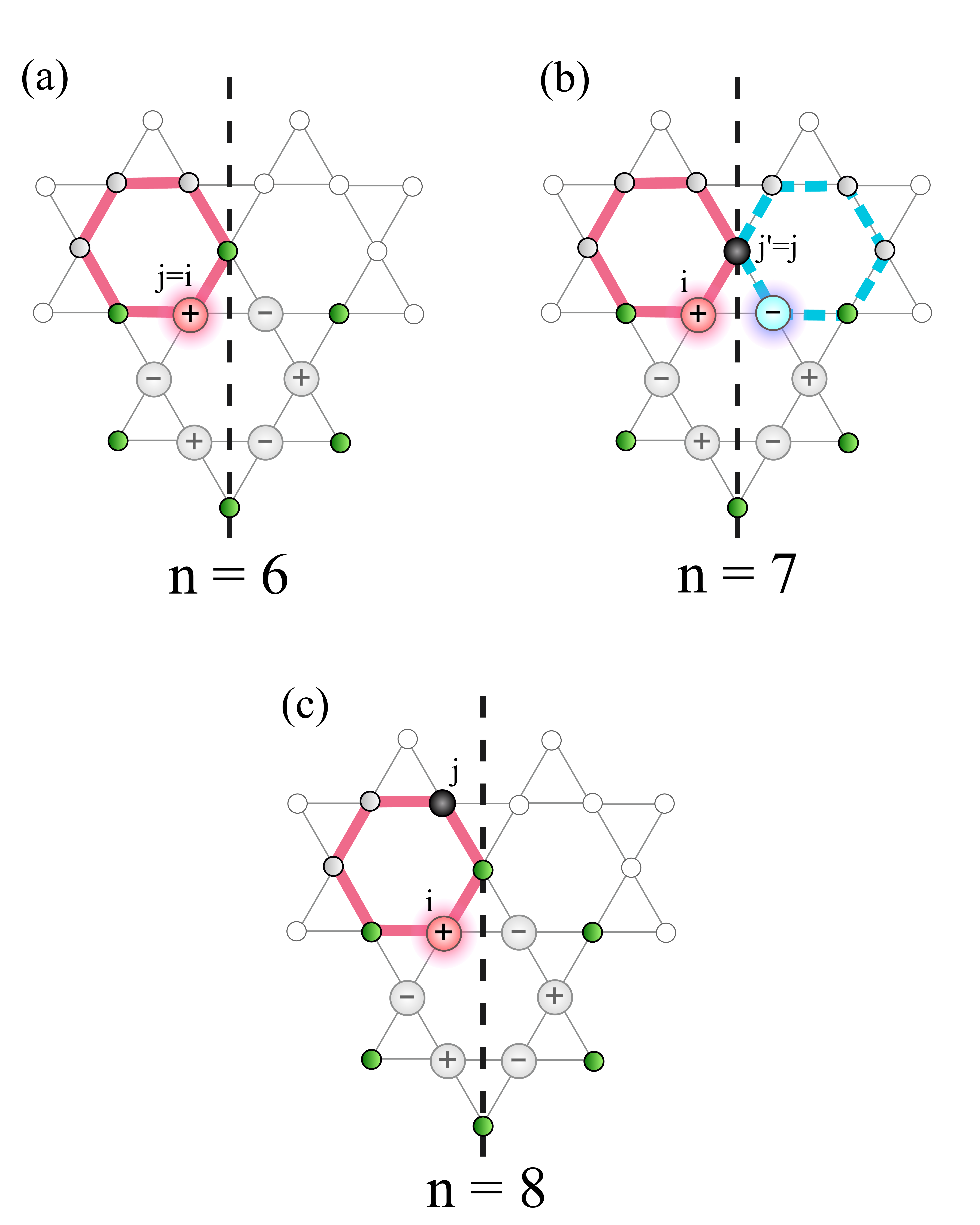}
	\caption{(a) Illustration of the only looped process in order $n=6$ as thick bonds. The particle in the inital and final state is located at the same site ($|i\rangle=|j\rangle$). Hence, this process does not conrtibute to the dispersion.
	(b) Hopping process contributing for orders $n\geq 7$. The corresponding counter process is represented by thick blue dashed bonds.
	(c) Hopping process contributing for orders $n\geq 8$. For this case any counter cluster has a different topology leading to the breakdown of the local mode $|\psi_{\rm loc}\rangle$.}
	\label{Fig:kagome_processes}
\end{figure}

The next crucial step is to identify the lowest pertubation order for which a hopping process exists which has at least one counter cluster with a different topology. In general those graphs are related to closed loops comprising two different $p^*_{ij}$ that do not lie on the same symmetry line as any counter cluster cuts such a loop. Clearly, the smallest nontrivial process with a loop is of order $n=6$ as shown in Fig.~\ref{Fig:kagome_processes}(a). In this order, the hopping process on this subcluster is always local ($i=j$) for the TFIM and hence does not contribute to the dispersion. For $n=7$ the only process with a loop, where the final state $|j\rangle$ is not part of $|\psi_{\rm loc}\rangle$ (and therefore could lead to a dispersion), is displayed in Fig.~\ref{Fig:kagome_processes}(b). But, this process has the special property that the site $j$ is located on a symmetry line of the lattice (one finds $j=j'$ for the counter cluster).  

In compliance with our results, the first hopping process leading to a breakdown of $|\psi_{\rm loc}\rangle$ is indeed of order $n=8$. Figure \ref{Fig:kagome_processes} depicts one such hopping process $t_{ij}^{C_{\rm sub},(8)}$ on the corresponding $C_{\rm sub}$. Besides this subcluster, there are also other processes leading to a dispersion in order $n=8$. All of them are related to non-trivial processes with loops.

Based on these geometrical observations for the kagome TFIM one finds the following criteria for the existence and the breakdown of local modes on a given lattice. Consider an arrangement of one-particle states located inside a restricted area $\mathcal{L}$. The sites enclosing this local area are denoted by $p^{*}_{rr'}$. It is clear that any subcluster leaving $\mathcal{L}$ has to pass through at least one $p^{*}_{rr'}$ which one can think of as a gate. In order to allow destructive interference every gate $p^*_{rr'}$ has to be connected to at least two sites belonging to $\mathcal{L}$. This restricted area provides a local mode up to a certain order $n$, if it allows for a superposition of one-particles states with alternating sign $\pm 1$ inside $\mathcal{L}$, such that all hopping processes $t_{ij}^{C_{\rm sub},(m)}$ on all subclusters $C_{\rm sub}$ for $m\leq n$ have a topologically equivalent counter cluster. For the triangular lattice, it turns out that there is no local arrangment that complies to the stated criteria.

General construction rules for localized magnon states are already well known from the intensive study in the context of macroscopic magnetization jumps \cite{schulenburg2002}. Our geometrical observations are fully consistent with the general rules stated in these works. 
Beyond these construction rules, we are able to identify single processes that lead to a dispersion which also allows to determine the exact order of locality on the basis of a quasi-particle picture.  The dispersion of a localized mode is affected by virtual processes that close a loop between different $p^{*}_{ij}$. The $p^{*}_{ij}$ are sites enclosing the local mode located in $\mathcal{L}$. Two sites $p^*_{ij}$ are considered as different if they do not share the same sites in $\mathcal{L}$.

\begin{figure}[htbp]
	\centering
		\includegraphics[width=0.6\columnwidth]{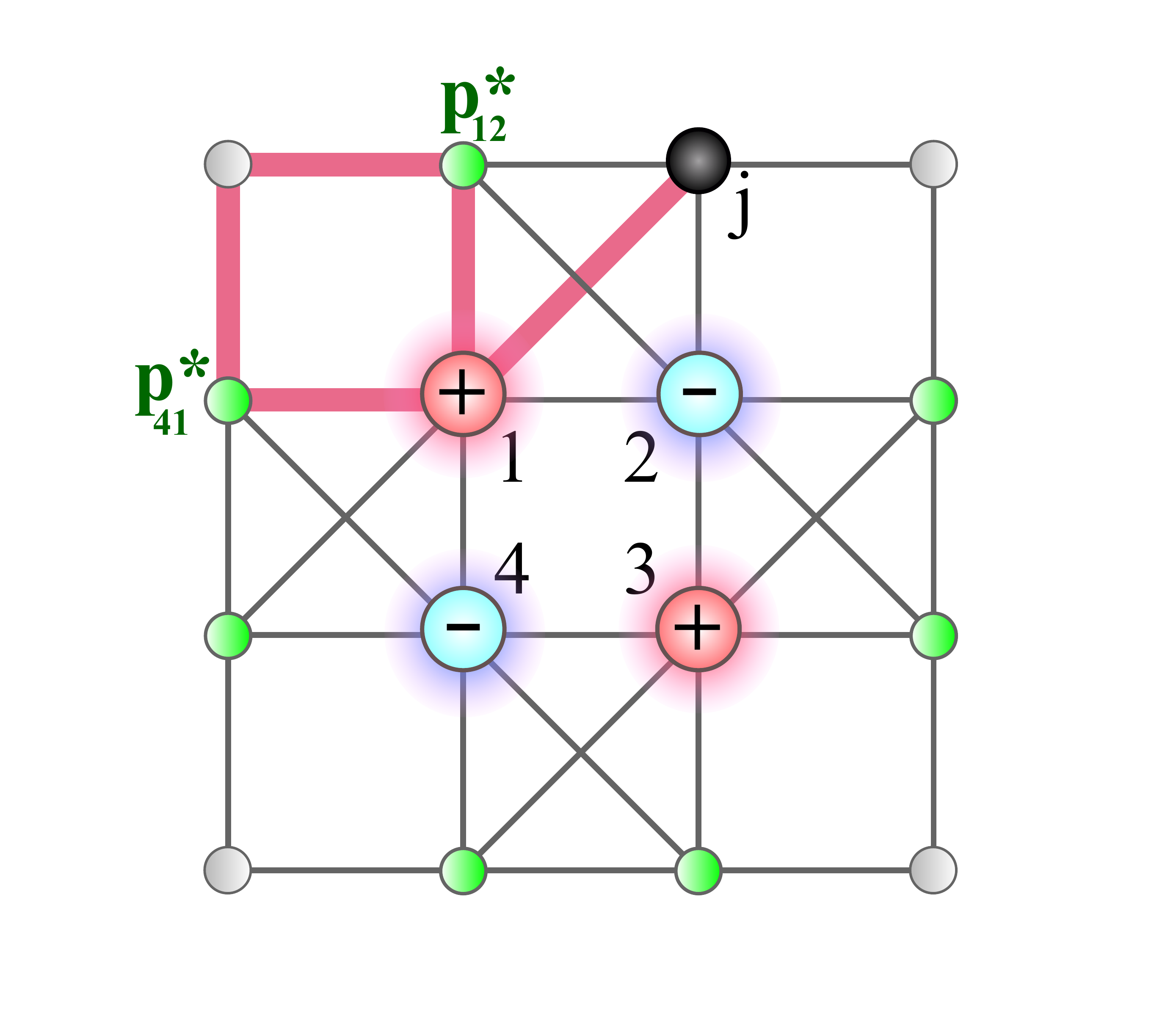}
	\caption{Localized mode $|\psi_{\rm loc}^{\rm sq}\rangle=\frac{1}{2}\left(|1\rangle-|2\rangle+|3\rangle-|4\rangle\right)$ on the square plaquette of the checkerboard lattice. The displayed hopping process from site $1$ to site $j$ involving 5 thick solid bonds leads to a dispersion as it closes a loop between the two different sites $p^{*}_{\rm 41}$ and $p^{*}_{\rm 12}$.}
	\label{Fig:checkerboard}
\end{figure}

Let us illustrate these principles by means of another TFIM on a frustrated lattice. For instance, consider the checkerboard lattice shown in Fig.~\ref{Fig:checkerboard}. Here the localized state  $|\psi_{\rm loc}^{\rm sq}\rangle$ is given by an arrangement of particles with alternating phase located at the sites of a square plaquette. The displayed process has a non-trivial loop containing two different sites $p^{*}_{ij}$ leading to a breakdown of $|\psi_{\rm loc}^{\rm sq}\rangle$. In the case of the TFIM the corresponding hopping process first appears in order $n=5$. Calculations for the checkerboard TFIM indeed show that the dispersion of the lowest one-particle mode is completely flat up to and including order $n=4$ in accordance with our predictions.

\section{Summary}\label{sec:summary}
We have studied the triangular and the kagome TFIM at zero temperature. Such highly-frustrated Ising models represent the simplest examples displaying an intriguing interplay of classical frustration and quantum fluctuations. Indeed, the tranverse magnetic field represents a singular perturbation on the extensive number of ground states present for the classical Ising problem. 

Generically, one expects an order by disorder scenario for such problems \cite{moessner2000,moessner2001}. The corresponding zero-temperature phase diagram then consists of two phases, an ordered one at low fields and a polarized phase at high fields separated by a quantum phase transition. Alternatively, a fascinating disorder by disorder scenario is possible. Here no symmetry-broken phase is realized and a polarized phase is present on the whole parameter axis. 

In this work we give strong evidence that the kagome TFIM displays a notable first instance of an disorder by disorder scenario in two dimensions. In contrast, the triangular TFIM falls in the more conventional class of systems showing an order by disorder.

Our findings are based essentially on high-order series expansions of the one-particle gap about the high-field limit. To this end a full linked graph expansion is performed using the pCUT method. For the triangular TFIM extrapolations of the one-particle gap give convincing evidence for a gap closing and therefore for a second-order quantum phase transition. The location of the critical point is in quantitative agreement with quantum Monte Carlo simulations \cite{moessner2003}. Furthermore, the obtained critical exponent is fully compatible with the suggested 3d XY universality class. 

In contrast, the extrapolation of the one-particle gap for the kagome TFIM displays no tendency to close. Most of the extrapolations suggest a finite gap of the order $\approx 0.2$ in units of $2h$ even in the low-field limit, which is consistent with first-order degenerate perturbation theory about the classical Ising antiferromagnet. The latter is done for finite systems yielding a gap $\approx 0.18$ in units of $2h$ in the thermodynamic limit. Put together, we have determined the one-particle gap on the whole parameter axis. Strong evidence is found that the low-field and high-field limit are adiabatically connected. Consequently, the kagome TFIM realizes a disorder by disorder scenario.

Let us also address similarities and differences between the impact of quantum fluctuations (in this work induced by the transverse field) and of thermal fluctuations on classical magnets on the kagome lattice. Whereas thermal fluctuations also do not induce order for Ising spins, the situation is different for the Heisenberg case. Here the high-temperature expansion has an analogous structure, in particular with regard to the momentum space degeneracy up to order eight in J/T \cite{harris1992}. There, up to Mermin-Wagner physics, dipolar, quadrupolar, and octupolar orders appear to be induced by thermal fluctuations \cite{huse1992,harris1992,chalker1992,ritchey1993,reimers1993,zhitomirsky2008,chern2012}.

The momentum independence of the low-energy band in the first seven orders of perturbation theory for the kagome TFIM can be traced back to the existence of local modes in the spatial domain. It is only in order eight that quantum fluctuations occur which do not interfere desctructively and therefore lead to the breakdown of the local modes. We have identified a mechanism for the (dis)appearance of this degeneracy which readily applies in a generalised setting. 

Our findings might inspire future research about similar questions in other classes of problems beyond transverse field Ising models. One fascinating idea would be to establish a link between the importance of flat bands in magnetic systems discussed in this work and the physics of fractional Chern insulators also based on the existence of flat bands on the one-particle level.

\section{Acknowledgements}
KPS acknowledges ESF and EuroHorcs for funding through his EURYI. RM thanks Premi Chandra, Sergei Isakov and Shivaji Sondhi for collaboration on closely related work. This work was in part supported by the Helmholtz Virtual Institute "New states of matter and their excitations".

\end{document}